\documentstyle[12pt]{article}
\catcode`\@=11
\long\def\@makefntext#1{ 
\protect\noindent \hbox to 3.2pt {\hskip-.9pt
$^{{\ninerm\@thefnmark}}$\hfil}#1\hfill} 

\def\thefootnote{\fnsymbol{footnote}}
\def\@makefnmark{\hbox to 0pt{$^{\@thefnmark}$\hss}}  

\def\ps@myheadings{\let\@mkboth\@gobbletwo
\def\@oddhead{\hbox{} 
\rightmark\hfil\ninerm\thepage}
\def\@oddfoot{}\def\@evenhead{\ninerm\thepage\hfil 
\leftmark\hbox{}}\def\@evenfoot{}
\def\sectionmark##1{}\def\subsectionmark##1{}}

\begin{document}

\newcommand{\symbolfootnote}{\renewcommand{\thefootnote}
{\fnsymbol{footnote}}}
\renewcommand{\thefootnote}{\fnsymbol{footnote}}
\newcommand{\alphfootnote}
{\setcounter{footnote}{0}
\renewcommand{\thefootnote}{\sevenrm\alph{footnote}}}

\newcounter{sectionc}\newcounter{subsectionc}\newcounter{subsubsectionc}
\renewcommand{\section}[1] {\vspace{0.6cm}\addtocounter{sectionc}{1}
\setcounter{subsectionc}{0}\setcounter{subsubsectionc}{0}\noindent
	{\bf\thesectionc. #1}\par\vspace{0.4cm}}
\renewcommand{\subsection}[1] {\vspace{0.6cm}\addtocounter{subsectionc}{1}
	\setcounter{subsubsectionc}{0}\noindent
	{\it\thesectionc.\thesubsectionc. #1}\par\vspace{0.4cm}}
\renewcommand{\subsubsection}[1]
{\vspace{0.6cm}\addtocounter{subsubsectionc}{1}
	\noindent {\rm\thesectionc.\thesubsectionc.\thesubsubsectionc.
	#1}\par\vspace{0.4cm}}
\newcommand{\nonumsection}[1] {\vspace{0.6cm}\noindent{\bf #1}
	\par\vspace{0.4cm}}

\newcounter{appendixc}
\newcounter{subappendixc}[appendixc]
\newcounter{subsubappendixc}[subappendixc]
\renewcommand{\thesubappendixc}{\Alph{appendixc}.\arabic{subappendixc}}
\renewcommand{\thesubsubappendixc}
	{\Alph{appendixc}.\arabic{subappendixc}.\arabic{subsubappendixc}}

\renewcommand{\appendix}[1] {\vspace{0.6cm}
        \refstepcounter{appendixc}
        \setcounter{figure}{0}
        \setcounter{table}{0}
        \setcounter{equation}{0}
        \renewcommand{\thefigure}{\Alph{appendixc}.\arabic{figure}}
        \renewcommand{\thetable}{\Alph{appendixc}.\arabic{table}}
        \renewcommand{\theappendixc}{\Alph{appendixc}}
        \renewcommand{\theequation}{\Alph{appendixc}.\arabic{equation}}
        \noindent{\bf Appendix \theappendixc #1}\par\vspace{0.4cm}}
\newcommand{\subappendix}[1] {\vspace{0.6cm}
        \refstepcounter{subappendixc}
        \noindent{\bf Appendix \thesubappendixc. #1}\par\vspace{0.4cm}}
\newcommand{\subsubappendix}[1] {\vspace{0.6cm}
        \refstepcounter{subsubappendixc}
        \noindent{\it Appendix \thesubsubappendixc. #1}
	\par\vspace{0.4cm}}

\def\abstracts#1{{
	\centering{\begin{minipage}{30pc}\tenrm\baselineskip=12pt\noindent
	\centerline{\tenrm ABSTRACT}\vspace{0.3cm}
	\parindent=0pt #1
	\end{minipage} }\par}}

\renewenvironment{thebibliography}[1]
        {\begin{list}{\arabic{enumi}.}
        {\usecounter{enumi}\setlength{\parsep}{0pt}
\setlength{\leftmargin 1.25cm}{\rightmargin 0pt}
\setlength{\leftmargin 0.52cm}{\rightmargin 0pt}
         \setlength{\itemsep}{0pt} \settowidth
        {\labelwidth}{#1.}\sloppy}}{\end{list}}

\topsep=0in\parsep=0in\itemsep=0in
\parindent=1.5pc

\newcounter{itemlistc}
\newcounter{romanlistc}
\newcounter{alphlistc}
\newcounter{arabiclistc}
\newenvironment{itemlist}
    	{\setcounter{itemlistc}{0}
	 \begin{list}{$\bullet$}
	{\usecounter{itemlistc}
	 \setlength{\parsep}{0pt}
	 \setlength{\itemsep}{0pt}}}{\end{list}}

\newenvironment{romanlist}
	{\setcounter{romanlistc}{0}
	 \begin{list}{$($\roman{romanlistc}$)$}
	{\usecounter{romanlistc}
	 \setlength{\parsep}{0pt}
	 \setlength{\itemsep}{0pt}}}{\end{list}}

\newenvironment{alphlist}
	{\setcounter{alphlistc}{0}
	 \begin{list}{$($\alph{alphlistc}$)$}
	{\usecounter{alphlistc}
	 \setlength{\parsep}{0pt}
	 \setlength{\itemsep}{0pt}}}{\end{list}}

\newenvironment{arabiclist}
	{\setcounter{arabiclistc}{0}
	 \begin{list}{\arabic{arabiclistc}}
	{\usecounter{arabiclistc}
	 \setlength{\parsep}{0pt}
	 \setlength{\itemsep}{0pt}}}{\end{list}}

\newcommand{\fcaption}[1]{
        \refstepcounter{figure}
        \setbox\@tempboxa = \hbox{\tenrm Fig.~\thefigure. #1}
        \ifdim \wd\@tempboxa > 6in
           {\begin{center}
        \parbox{6in}{\tenrm\baselineskip=12pt Fig.~\thefigure. #1 }
            \end{center}}
        \else
             {\begin{center}
             {\tenrm Fig.~\thefigure. #1}
              \end{center}}
        \fi}

\newcommand{\tcaption}[1]{
        \refstepcounter{table}
        \setbox\@tempboxa = \hbox{\tenrm Table~\thetable. #1}
        \ifdim \wd\@tempboxa > 6in
           {\begin{center}
        \parbox{6in}{\tenrm\baselineskip=12pt Table~\thetable. #1 }
            \end{center}}
        \else
             {\begin{center}
             {\tenrm Table~\thetable. #1}
              \end{center}}
        \fi}

%
%
%
\font\twelvebf=cmbx10 scaled\magstep 1
\font\twelverm=cmr10 scaled\magstep 1
\font\twelveit=cmti10 scaled\magstep 1
\font\elevenbfit=cmbxti10 scaled\magstephalf
\font\elevenbf=cmbx10 scaled\magstephalf
\font\elevenrm=cmr10 scaled\magstephalf
\font\elevenit=cmti10 scaled\magstephalf
\font\bfit=cmbxti10
\font\tenbf=cmbx10
\font\tenrm=cmr10
\font\tenit=cmti10
\font\ninebf=cmbx9
\font\ninerm=cmr9
\font\nineit=cmti9
\font\eightbf=cmbx8
\font\eightrm=cmr8
\font\eightit=cmti8

\newcommand{\ra}{\rightarrow}
\newcommand{\st}{\stackrel}
\newcommand{\se}{\simeq}
\newcommand{\lr}{\leftrightarrow}
\newcommand{\ep}{\epsilon}
\newcommand{\bi}{\bibitem}
\newcommand{\ds}{\displaystyle}
\newcommand{\bq}{\begin {eqnarray} & \ds}
\newcommand{\eq}{ & \end {eqnarray}}
\newcommand{\ov}{\overline}
\newcommand{\tr}{& \nonumber \\ & \ds}

\begin{center}{\hspace {8 cm} hep-ph/ 9503208}\end{center}
\begin{center}{\hspace {7.8 cm} Budker INP 95-18}\end{center}
\* \vspace{1cm}
\begin{center}
 {\bf CALCULATION OF THE D AND B MESON LIFETIMES\\

 AND THE UNITARITY TRIANGLE PARAMETERS}\\
\vspace{1.5cm}
 {\*\hspace{-1.5cm}} VICTOR \,\,CHERNYAK  \\
\vspace{1 cm}
{\bf \*\hspace{-1.cm} Budker Institute of Nuclear Physics, \,\\
\*\hspace{-1.5cm} 630090\, Novosibirsk-90}\\{\*\hspace{-1.5cm} and}
\\ {\bf \*\hspace {-1.5cm} Novosibirsk State University} \end{center}
\vspace{1cm}

 {\bf \abstract}

Using the expansions of the heavy meson decay widths in the heavy quark mass
and QCD sum rules for estimates of corresponding matrix elements,\, we
calculate the $D^{\pm,o,s}$ and $B^{\pm,o,s}$ meson lifetimes.
The results for D mesons are in a reasonable agreement with the
data,\, while it is predicted: $[\Gamma (B_d)-\Gamma (B^\pm)]/\Gamma_B\se
4\%\,$ (and the lifetime difference of the $B_d$ and $B_s$ mesons is even
smaller);\,
$[\Gamma(B_s^{short})-\Gamma(B_s^{long})]/{\ov \Gamma}(B_s)\se 8\%\,.$
The role of the weak annihilation and Pauli interference
contributions to the lifetime differences is described in detail. In the
course of self-consistent calculations the values of many parameters crucial
for calculations with charmed and beauty mesons are found. In particular,\,
the quark pole masses are: $M_c\se 1.65\,GeV,\,\, M_b\se
5.04\,GeV\,,$ and the decay constants are: $f_D(M_c)\se
165\,MeV\,,\,\,f_B(M_b)\se 120\,MeV\,$.  It is also shown that the
nonfactorizable corrections to the $B-{\bar B}$ mixing are large,\,
$B_B(M_b)\se (1-18\%)\,.$ The values of the unitarity triangle parameters are
found which are consistent with these results and the data available (except
for the NA31 result for the $\epsilon ^{\prime}/\epsilon$ which is too large):
$|V_{cb}|\se 4.2\cdot 10^{-2}\,,\, |V_{td}|\se 1.3\cdot 10^{-2}\,,\,
|V_{ub}/V_{cb}|\se 0.10\,,\, \{\,A\se 0.86\,,\,\,\rho \se -0.40\,,\,\, \eta
\se 0.20\,\}.$
\vspace{1cm}

\begin{center} Talk given at the WE-Heraeus Seminar "Heavy Quark Physics",\\
              Bad Honnef, Germany, December 94 \end{center}

\renewcommand{\thefootnote}{\arabic{footnote}}

\newpage
 \section {Introduction}

 Experimentally measured properties of heavy mesons offer the possibility
to find out values of parameters which are of fundamental importance for
the Standard Model: $M_c,\,M_b,\,V_{cb},\,V_{ub},\,V_{td}$ (also $f_D,\,
f_B,$\, etc). Below the attempt is described \cite {Cher1}
of self-consistent calculation
of all the above parameters using the available experimental data on the
D and B mesons. The theoretical methods used are: expansions in the heavy
quark mass for obtaining the effective Lagrangians and QCD sum rules for
estimates of corresponding matrix elements.  The used scheme of calculations
looks as follows:\\
 Experiment $\Gamma(D\ra e\nu+X):\,\ra $ determination of $M_c$;\\
 $M_c+$ Heavy meson mass formulae: $\ra$ calculation of $M_b$;\\
 $M_c,\,M_b+$ QCD sum rules: $\ra$ calculation of the decay constants
$f_D,\, f_B$ and the matrix elements $<M|\,L_{eff}\,|M>$;\\
 Calculation of the $D^{\pm,o,s}$ meson lifetimes and predictions for the
$B^{\pm,o,s}$ meson lifetime differences and branching ratios;\\
 Calculation of the matrix element: $<{\bar B}^o|\,L_{eff}^{mixing}\,
|B^o>,\,$ prediction of\\ $\Gamma (B_s^{short})-\Gamma (B_s^{long})$.\\
 Experiment $\Gamma(B\ra e\nu+X):\,\ra $ determination of $|V_{cb}|$;\\
 Experiment $\Delta m_d\,:\,\ra$ determination of $|V_{td}|$;\\
 Experiment on $\epsilon_K:\,\ra$ determination of $\rho,\,\eta,\,V_{ub},$\,
estimates of $\epsilon^{\prime}/\epsilon$,\, predictions for CP-violation in
B-decays, etc.

\section {The history}

It is a long standing challenge for theory to calculate the D and B meson
decay widths. On the qualitative side,\, two mechanisms were invoked to
explain the pattern of the D meson lifetime differences: weak annihilation
(WA) \cite {FM},\, and Pauli interference (PI)
\cite{Gub},\,\cite{VS}. As for WA,\, it was expected
that because an admixture of the wave function component with an additional
gluon or the emission of a perturbative gluon,\, both
remove a suppression due to helicity conservation (which leads to $Br\,
(\pi\ra e\,\nu)/Br\,(\pi\ra \mu\,\nu)\sim 10^{-4})$,\, the $D^o$ meson decay
width is enhanced. On the other hand,\, it was expected that the
destructive PI of two d-quarks (the spectator
and those from a final state) will suppress the $D^{\pm}$ meson
decay width.  As for WA,\, there were no reliable calculations at all. For
PI,\, simple minded estimates (see sect.7) give too large an effect which
results in a negative $D^{\pm}$ decay width.

For B mesons,\, it was clear qualitatively that all the above effects
which are of a pre-asymptotic nature and die off at $M_Q\ra \infty$,\, will be
less important. However,\, because the patern of the D meson lifetime
differences was not really explained and well understood,\, this prevented to
obtain reliable estimates of the B meson lifetime differences,\, and only
order of magnitude estimates were really available: $[\delta \Gamma(B)/\Gamma_
B]:[\delta \Gamma(D)/\Gamma_D]\sim O(f_{B}^{2}M_{c}^{2}/f_{D}^2M_{B}^{2})
\sim O(10^{-1})$.

Moreover,\, as for WA contributions through perturbative gluon
emission (which is formally a leading correction $\sim
O(\Lambda_{QCD}/M_Q)$ to the deacay width
and was expected before to be potentially the most important),\, it has
been emphasized recently \cite{BU}
that such contributions are of no help at all
because,\, being large (at least formally at $M_Q \ra \infty$) term by term,\,
they cancel completely in the inclusive widths,\, both $O(1/M_Q)$ and $O(
1/M_Q^2)$ terms.
It will be shown below (see sects.8-10) that,\, nevertheless,\, there
are important WA contributions but on the nonperturbative level.

Considerable progress has been achieved recently in applications of the
operator product expansion to the calculation of the heavy meson decay width.
In particular,\, it was shown that there are no $O(\Lambda_{QCD}/M_Q)$
corrections to the Born term and first nonperturbative corrections
$O(\Lambda_{QCD}^2/M_Q ^2)$ were calculated explicitly
\cite{BUV1},\,\cite{BS1}. However,\, these contributions are all nonvalence
and so have nothing to do with lifetime differences. They are important
however for the calculation of the absolute decay rates. There appeared a
number of papers where these results (neglecting the four-fermion operator
contributions) were applied to determine the values of the quark masses,\,
$M_c$ and $M_b$,\, and $|V_{cb}|$ \cite{LS},\,\cite{Big},\,\cite{LN}.
However,\, because the four-fermion operator contributions
 are of crucial importance for the D mesons and
lead to $\tau(D^+)/\tau(D^o)\se 2.5$,\, the real accuracy of the above results
remained unclear.  \footnote{ In fact,\, the authors of \cite{Big} renounced
their results on $M_c$ and $M_b$ and used $M_c=1.35-1.40 \,GeV,\,M_b=4.8\,GeV$
in their later articles \cite{BBSUV}.}

 \section {\bf The heavy quark masses}

 As most of calculations with heavy quarks are highly sensitive to precise
 values of their masses,\, it is of prime importance to know these as
 precisely as possible.

 Below the pole masses,\, $M_c$ and $M_b,\,$ are used for the charm and bottom
 quarks as most convenient gauge and renormalization scheme independent
 quantities. Besides, the most convenient expansion parameter of the Heavy
 Quark Effective Theory (HQET)\cite{IW} is just the quark pole mass. On the
 other hand,\, because the pole mass receives contributions from the
 infrared region,\, the perturbative series connecting it with the current
 mass (say,\, ${\ov {MS}}$-mass) is divergent due to renormalon
 effects \cite{BSUV}:
\bq \*\hspace{-0.3cm} M_Q={\ov M_Q}\left\{1+\frac{4}{3}\frac{\alpha_s
(M_Q)}{\pi}+1.56\, b_o \left(\frac{\alpha_s(M_Q)}{\pi}\right)^2+
\cdots \right\} \equiv \tr \equiv {\ov M_Q} \left\{1+\frac{4}{3}
 \frac{\alpha_s(M_Q)}{\pi} \kappa^{(m)}_Q\right\}.\eq
 Because the series in Eq.(1) is divergent,\, the result contains an
 ambiguity of order $O(\Lambda_{QCD})$ and requires for a concrete
definition.  It is possible,\, for instance,\, to cut out the series at the
 optimal number of terms: $n\se 2\pi/(b_o\alpha_s)$.  In what follows we
 choose the definition of the quark pole mass which looks most natural,\, i.e.
 the real part of the analytically continued Borel transform (\,Re[BT]\,) of
the above series\,(see i.e. \cite{FKW}).  This was calculated in \cite{BB}
(in the improved leading order approximation,\, ILO,\, i.e. by replacing
constant $\alpha_s$ by running $\alpha_s$ in one loop corrections),\, so that
the result for $\kappa^{(m)}_Q$ in Eq.(1) is known.

 To illustrate the characteristic numbers,\, let us take (see below): $M_c=
(1.65\pm 0.05)\,GeV,\,M_b=(5.04\pm 0.05)\,GeV$. Then ($\alpha_s$ is in the
${\ov {MS}}$-scheme: $\,\alpha_s(M_c)=0.310,\, \alpha_s(M_b)=0.204$)\,:
$\kappa^{(m)}_c=1.75,\, \kappa^{(m)}_b=2.10$\, \cite{BB}\,:
\bq M_c={\ov M}_c\left
[1+\frac{4}{3}\frac{\alpha_s(M_c)}{\pi}\left (1+1.04+\cdots\right )\right
]= \tr =
{\ov M}_c \left [1+\frac{4}{3}\frac{\alpha_s(M_c)}{\pi}1.75\right ]=
1.23\,{\ov M}_c\,,\eq
\bq M_b={\ov M}_b\left (1+\frac{4}{3}\frac{\alpha_s(M_b)}{\pi}\left (1+
0.62+\cdots \right )\right ]= \tr =
{\ov M}_b\left [1+\frac{4}{3}\frac{\alpha_s(M_b)}{\pi}2.10\right ]=
1.18\,{\ov M}_b\,.\eq
Therefore,\, the value $M_c=1.65\,GeV$ found below from the data on $\Gamma
(D\ra e\nu+X)$ leads to: ${\ov M}_c=1.34\,GeV$,\, which is $\se 70\,MeV$
above the value ${\ov M}_c=1.27\,GeV$ obtained long ago \cite{Nov} from the
charmonium sum rules. The value $M_b=5.04\,GeV$ obtained below
from $M_c=1.65\,GeV$ and HQET mass formulae leads to: ${\ov M}_b=4.27\,GeV$,\,
in  good agreement with the
value ${\ov M}_b=4.25\,GeV$ obtained in \cite {Nov}
from the upsilonium sum rules.

It seems natural to expect that the accuracy of ${\ov M}_b$ obtained
from the upsilonium sum rules is better in comparison with ${\ov M}_c$
obtained from the charmonium. So,\, supposing the value ${\ov M}_b=4.27\,GeV$
is sufficiently
accurate (say,\, within a several tens MeV),\, we can then turn around the
chain of reasoning. Starting from this value we obtain $M_b=5.04\,GeV$ from
Eq.(3), \footnote{
\,It seems,\, this value of $M_b$ is stable even on account
of $O(\alpha_s^2 )$ corrections. On the one hand,\, there appeared
(preliminary) results \cite {Broad} on the $O(\alpha_s^2)$-corrections to the
sum rules . They are positive and enter with significant coefficients. So,\,
they will increase slightly the above value ${\ov M}_b=4.25\,GeV.$ On the
other hand,\, the genuine $O(\alpha_s^2)$-correction in Eq.(3) is known and is
negative,\, so that the transition coefficient between ${\ov M}_b$ and $M_b$
is slightly less than 1.18\,. These two effects tend to compensate each
other,\, so that the value $M_b=5.04\,GeV$ has real chanses to be sufficiently
precise.}
\,\,and $M_c=1.65\,GeV$ from the HQET mass formulae,\, with each
step having a several tens MeV typical accuracy.

\section { $D^{+}\ra l\nu+X$. \quad Determination of $M_c$.}

 On account of radiative,\, $O(\Lambda_{QCD}^2/M_c^2)$ and four-fermion
 operator ($\,O(\Lambda_{QCD}^3/ M_c^3)$ corrections,\, the expression for
$\Gamma_{sl}$ can be represented in the form \cite{BUV1}:
 \bq \Gamma_{sl}(D^{+})= \Gamma^{sl}_{Born}\,I_{rad}\left \{\,z_o\,N_c
\left [1-\frac{2}{N_c}\frac{z_1}{z_o}\frac{\mu^2_G}{M_c^2}\right ]
   +\delta_{lept}^{(c)}\pm \Delta\right \}\,, \eq
   \bq \Gamma_{Born}^{sl}=\frac{G_F^2\,M_c^5}{192\,\pi^3}\,,\quad
  N_c=\frac{<D|\,{\bar c}\,c\,|D>}{2\,M_D}\se \left (1+\frac{\mu_G^2-
 \langle {\bf p}^2\rangle }{2M_c^2}\right )\,,\eq
 \bq \langle{\bf p}^2\rangle=\frac{<D|{\bar c}(i{\bf D})^2c|D>}{2M_D},\quad
   \mu_G^2=\frac{<D|{\bar c}\frac{i}{2}g_s\sigma_{\mu\,\nu}G_{\mu\,\nu}c
   |D>}{2M_D}\,,\eq
  \bq  I_{rad}\se \left [1-\frac{2}{3}\,\frac{\alpha_{s}(M_c)}{\pi}\,
   f_{o}\,\kappa^{(w)}_c \right ]\,,\eq
  where $z_o,\,z_1,\,f_o$ are known functions of $m_s/M_c$,\,
  $\kappa^{(w)}_c$ is the analog of $\kappa^{(m)}_c$ in Eq.(1) and
  $\delta_{lept}^{(c)}$ in Eq.(4) is the contribution of the
  four-fermion operators.

    We use below\,
   \footnote{
   Our definition of $\langle {\bf p}^2\rangle $ (and,\, analogously,\, ${\ov
 \Lambda}$ and others) differs from the cut off dependent
   $\langle {\bf p}^2\rangle_\mu$
   used by I.Bigi et.al.  If one represents this last (at $\mu\gg
   \Lambda_{QCD}$) as:  \bq \langle {\bf
   p}^2\rangle_\mu=C_o\frac{\alpha_s}{\pi}\mu^2+\cdots
   +C_1\Lambda_{QCD}^2+C_2\frac{\Lambda_{QCD}^3}{\mu}+\cdots\,,\eq
   then our $\langle {\bf p}^2\rangle $ corresponds to the cut off independent
   term $C_1\Lambda_{QCD}^2$. We prefer this definition because it selects the
   universal number,\, while all $\mu$-dependent terms will be canceled
   finally in observable quantities like decay widths. On the other hand,\,
   with our definition some useful inequalities like
   $\langle {\bf p}^2\rangle_{\mu}\geq
   (\mu_G^2)_\mu$ will be lost,\, in general.} :\,\,
   \footnote{
   From our viewpoint,\, the value: $\langle {\bf p}^2\rangle \se
   0.5-0.6\,GeV^2$ obtained in \cite{PBB} is overestimated. Let us recall
   \cite {Ioffe} that the mean value of the vacuum quark 4-momentum squared
   is:  $\langle -k_{\mu}^2\rangle_o=4/3\, \langle{\bf k}^2\rangle_o\se
   0.4\,GeV^2,\,$ and the quarks in the pion have their momenta somewhat less
   on the average than the vacuum quarks \cite{CZ5}.  Let us point out also
   that $\langle {\bf p}^2\rangle $ enters here to $N_c$
   only and plays no essential role.}
   \bq m_s\se 150\,MeV;\, \langle {\bf p}^2\rangle \se 0.25\,GeV^2;\,\,
   \mu_G^2\se \frac{3}{2}M_c\left (M_{D^*}-M_D \right )\se 0.35\,GeV^2.\eq

  The value of $M_c$ can be determined from a comparison
   of Eq.(4) with the experimental value:
  $\Gamma_{sl}(D^{+}\ra l\nu+X)=(1.08\pm 0.06)\cdot 10^{-13}\,GeV\,.$
  Because the dependence of
  $\Gamma_{sl}$ on $M_c$ is highly nonlinear,\, it is more convenient to
  proceed in an opposite way.  Namely,\, let us show that Eq.(4) reproduces
  the experimental value at $M_c\se 1.65 \,GeV$. We have:
   \bq \alpha_{s}(\,M_{c}^{2}\,)\se 0.310\,,\quad f_o\se 3.25\,,
  \quad N_c\se 1.02\,, \quad
  \Gamma_{Born}^{sl}\se 2.80\cdot 10^{-13}\,GeV\,,\eq
  \bq \Gamma_{sl}(D^+)\se \Gamma^{sl}_{Born}\,I_{rad}\left
  \{0.71+\delta_{lept}^ {(c)} \pm \Delta\right \}.\eq
  As for $\kappa^{(w)}_c$,\,
   the two loop radiative correction $O(b_o\alpha^2_s)$ was calculated in
  \cite{LSW}:  \bq I_{rad}=\left
 [1-\frac{2}{3}\frac{\alpha_s(M_c)}{\pi}f_o\left (1+1.03 +\cdots\right )
 \right ]\,.\eq
 Comparing with Eq.(2),\, it is seen that both series follow the same
 pattern.  Besides,\, because the leading renormalon is the same in both
 series,\, it seems clear that $\kappa^{(w)}_c$ will be close to
 $\kappa^{(m)}_c$.  \footnote{\, This is supported also by the examples
 considered in \cite{BB}.} {\,\,\,Therefore,}\, we estimate:
 $\kappa^{(w)}_c\se 1.75,\,$ so that $I_{rad}\se 0.626.$

The main contributions to $\delta_{lept}^{(c)}$ in Eq.(4) originate from
the figs.1a,b diagrams which show the matrix elements of the four-fermion
operators in the effective Lagrangian \cite{Cher1}:
\bq \left \{\delta^{(c)}_{lept}\right \}_{figs.1a,1b}\se \left \{-8.5\%\right
\}_{fig.1a}+ \left \{-3.5\%\right \}_{fig.1b}\se -12\%\,.\eq

There are also other small contributions to $\delta_{lept}^{(c)}$,\, the
typical one is from the four-fermion operator hidden in the Born operator
${\bar c}p^4{\hat p}c$,\, fig.1c
\footnote{
Our result here differs by the
factor 1/2 from those obtained in \cite{BDS}.}\,\,:
\bq \left \{\delta_{lept}^{(c)}\right \}_{fig.1c}=-\frac{2\pi\,\alpha_s}
{M_D\,M_c^3}<D|{\bar c}(0)\gamma_\mu(1+\gamma_5)\frac{\lambda^a}{2}c(0)\cdot
J^a_{\mu}(0)|D>\se 1\%\,.\eq
Estimates show that the typical value of next
corrections (denoted by $\pm \Delta$ in Eq.(4)) is a few per cent. To be
   safe,\, we take: $\Delta= 6\%.$ So\,:
\bq \Gamma_{sl}(D^+)\se 0.375\,\Gamma^{sl}_{Born}\left (1\pm 10\%\right )
\se \left (1.05\pm 0.10\right ) \cdot 10^{-13}\,GeV\,, \eq
in agreement with data. Because the decay width
is highly sensitive to the precise value of $M_c$,\,
\footnote{
For instance,\, with $M_c=1.55\,GeV$ the calculated $\Gamma_{sl}(D)$ will
be more than $40\%$ smaller.} {\,\,\,} this later is tightly constrained:
\footnote{
Really,\, higher order corrections are arranged in Eq.(4) in such a way to
obtain a minimal possible value of $M_c$. Therefore,\, $M_c=1.65\,GeV$ is
rather a lower bound.}
\bq M_c=\left (1.65\pm 0.05\right )\,GeV\,.\eq
We were going in this section into calculation details as the precise value
of $M_c$ is the base of all further calculations,\, and because the above
value of $M_c$ is essentially higher than the commonly accepted at present
values $M_c=1.35-1.45\,GeV$.

\section { Mass formulae: $M_b$ and ${\ov \Lambda}$.}

The HQET mass formulae look as (\,${\ov M}=(M_P+3M_V)/4\,$):
\bq M_b-M_c={\ov M}_B-{\ov M}_D+\left [\,\frac{\langle {\bf p}^2\rangle }
{2M_c}-\frac {\langle {\bf p}^2\rangle}{2M_b}\,\right ]+O\left (\frac
{\Lambda_{QCD}^3}{M_c^2}\right )\,,\eq
\bq M_B=M_b+{\ov \Lambda}+\frac{\langle {\bf p}^2\rangle-\mu_G^2}{2M_b}+
O\left (\frac{\Lambda_{QCD}^3}{M_b^2}\right )\,,\eq
The expected accuracy of Eq.(17) is a several tens MeV,
\footnote{
\,Supposing that $\langle {\bf p}^2\rangle $ is within the reasonable
interval,\, say:  $0.25\pm 0.05\,GeV^2\,.$} \, and even better in Eq.(18). We
obtain therefore (with the accuracy $\pm 50\,MeV$,\, determined mainly by the
uncertainty in $M_c$:
\bq M_b=5.04\,GeV\,,\quad {\ov \Lambda}=250\,MeV\,.\eq
These results differ significantly from the widely used values: $M_b=
4.8\,GeV,\, {\ov \Lambda}=500\,MeV$.

\section {${\bf f_D}\,$ and $ {\bf f_B}\,$.}

The knowledge of precise values of the decay constants $f_D$ and $f_B$
is of crucial importance for many calculations with the D and B mesons
(analogously to $f_{\pi}$ for the pion). The calculated values of $f_D,\,
f_B$ are highly sensitive to the precise values of  $M_c,\,M_b$ (more
precisely,\, to $M_D-M_c\se M_B-M_b\se {\ov \Lambda}$),\, and increase
quickly with increasing $\ov \Lambda$.

 The QCD sum rules for the chiral current correlator which is a difference of
the pseudoscalar and scalar current correlators,\, posses the advantage of
being protected,\, in the chiral limit,\, against pure perturbative
contributions (which are poorly controlable in separate correlators),\, and
having no significant loop corrections to nonperturbative contributions.
Using them and the above values of $M_c,\, M_b$,\, we obtain \cite{Cher1}:
\bq f_D(M_c)\se 165\,MeV\,,\quad f_B(M_b)\se 120\,MeV\,,\eq
(with the expected accuracy about $10\%$,\, which is always difficult to
estimate reliably when dealing with the QCD sum rules).

While the above value of $f_D$ is only slightly below the widely accepted
at present value $\se 180\,MeV$,\, the above value of $f_B$ is much
smaller than the widely used values $\se 180-200\,MeV$.
\footnote{
As for the QCD sum rule calculations,\, the large value of $f_B$ originates
mainly from using small values of $M_b$ (i.e. ${\ov \Lambda}\se 500-600\,
MeV$). As for the lattice calculations,\, the predictions for $f_B$
decrease with time,\, starting from $\se 250-300\,MeV$ 1-2 years
ago and reaching now $148\pm 20\,MeV$ in the latest paper \cite{Bern}.}

\section{Difficulties with the naive lifetime estimates.}

On account of the Born term,\, the leading radiative and $O(\Lambda_{
QCD}^2/M_c^2)$ corrections (all nonvalence),\, the D meson hadronic
width is ($\Gamma_{Born}=G_F^2M_c^2/64\pi^3$):
\footnote{\, Unlike the semileptonic width,\, there are two $\mu^2_G/M_c^2$
corrections,\, each one $\se 30\%$ but they cancel strongly each other
 \cite{BUV1}.}
\bq \Gamma_{nl}^o \se \Gamma_{Born}\,\bigl [\,1.50\, \bigr
]_{rad}\,z_o\, N_c\, \bigl [\,1.07\, \bigr ]_{\mu_G}\se
1.54\,\Gamma_{Born}\,.\eq

At the level $O(\Lambda_{QCD}^3/M_c^3)$ there appear first valence
(and additional nonvalence) contributions to the decay widths
originating from the four-fermion operators. The effective Lagrangian
(normalized at $\mu_o^2=0.5\,GeV^2$) has the form \cite{Cher1}:
\footnote{\,\, Only the most important terms are shown.}
   \bq L^{c}_{eff}\se\frac{G_F^2}{2\,\pi}\,\left
   \{\,{\bar \lambda}^2\,g_{\mu\,\nu}\,L^{d}_{\mu\,\nu}+\lambda^2\,
   t_{\mu\,\nu}(\lambda)\,L^{u}_{\mu\,\nu}+L_{PNV}+\cdots \right \}\,,\eq
   \bq L^{d}_{\mu\,\nu}=\left \{-1.1\,
   (\,{\bar c}\,\Gamma_{\mu}\,d\,)\,({\,\bar d}\,\Gamma_{\nu}\,c\,)+
   4.0\,({\,\bar c}\,\Gamma_{\mu}\frac{\lambda^a}{2}\,d\,)\,(\,{\bar d}\,
   \Gamma_{\nu}\frac{\lambda^a}{2}\,c\,)\,\right \}\,, \eq
   \bq L^{u}_{\mu\,\nu}= 3.3\,(\,{\bar c}\,\Gamma_{\mu}\frac
   {\lambda^a}{2}\,u)\,({\bar u}\,
   \Gamma_{\nu}\frac{\lambda^a}{2}\,c\,)\,, \quad
   t_{\mu\,\nu}(\lambda)=\frac{1}{3}\left ( \frac{\lambda_\mu\,\lambda_\nu}
   {\lambda^2}-g_{\mu\,\nu}\right )\,,\eq
   where $\lambda$ (or ${\bar \lambda}\,$) is the total 4-momentum of the
   integrated quark pair. It can be read off from each diagram in fig.2
   and differ from $P_c$ by the spectator quark momenta. The term $L_{
   PNV}$ is nonvalence and originates from the diagram in fig.1a.

 Let us try now to obtain the estimate of $<D^+|L_{eff}|D^+>$ using the
 factorization approximation and neglecting the spectator quark momenta in
 comparison with $M_c$ \cite{VS},\,\cite{BS2}.  Then,\, $L^u$ and the
 second term in $L^d$ give zero contributions and:
 \bq \Delta \Gamma_{factor}(D^{+})\se \left [ -1.1\cdot
 16\,\pi^2\,\frac{f_D^2(\mu_o)\,M_D}{M_c^3}\right ]\,\Gamma_{Born}\se -1.50\,
 \Gamma_{Born}\,.\eq
 The term $L_{PNV}$ in Eq.(22) adds $\se -0.15\,\Gamma_{Born}$\, \cite
 {Cher1},\, so that we obtain for the hadronic width of $D^+$:
 \bq \Gamma_{nl}(D^+)\se (1.54-1.50-0.15)\,\Gamma_{Born}\se -0.11\,
 \Gamma_{Born}\,,\eq
 which does not make much sense. It is clear that the above
 approximations are too poor and some estimates are essentially wrong.

 \section {Non-factorizable contributions,\,
    ${\bf \lambda}$ and ${\bf {\ov \lambda.}}$}

 It is seen from Eqs.(22)-(24) that (internally) coloured operators whose
matrix elements are zero in the factorization approximation,\, enter $L_{eff}$
with much larger coefficients. So,\, even if their matrix element are
suppressed,\, they may be of importance. This is really the case. For
instance,\, the most important weak annihilation non-factorizable
contributions shown in fig.2 were estimated using the QCD sum rules
\cite{Cher1}:  \bq <D(p)|({\,\bar
c}\,\Gamma_{\mu}\frac{\lambda^a}{2}q\,)\,(\,{\bar q}
   \,\Gamma_{\nu}\frac{\lambda^a}{2}\,c\,)|D(p)>_{fig.2a,b}\,\,\se \frac{1}
{15} f^2_D M_D^2 \left (\frac{p_\mu\,p_\nu}{p^2}-g_{\mu\,\nu}\right )\,,\eq
 \bq <B(p)|({\,\bar b}\,\Gamma_{\mu}\frac{\lambda^a}{2}q\,)\,(\,{\bar q}
  \,\Gamma_{\nu}\frac{\lambda^a}{2}\,b\,)|B(p)>_{fig.2a,b}\,\,\se \frac{1}{20}
 f^2_B M_B^2 \left (\frac{p_\mu\,p_\nu}{p^2}-g_{\mu\,\nu}\right )\,,\eq
 (the contribution due to the fig.3 diagram is smaller,\, the fig.2c
contribution is negligibly small).
Comparing with the factorizable contribution:
 \bq <D(p)|({\,\bar c}\,\Gamma_\mu\,q\,)\,(\,{\bar q}
   \,\Gamma_\nu\,c\,)|D(p)>\se f^2_D M_D^2\left (\,\frac{p_\mu\,p_\nu}
 {p^2}\,\right )\,, \eq
 we see that the factorization approximation works very well,\,
even for the D mesons.

 But there is one more factor which suppresses heavily the factorizable
 contributions in the D meson case: the phase space of the integrated
 quark pair is much smaller for the crossed contributions (fig.2b) in
   comparison with the direct ones (fig.2a). I.e.,\, the characteristic
   value of $\lambda^2$ in Eq.(22)
 is much larger than those of ${\ov \lambda}^2$. Really
 : $\lambda\se P_c+k=P_D,\,$ while ${\ov \lambda}\se P_D-k_1-k_2$,\,
   where $k_1,\,k_2$ are the spectator quark momenta. It can be shown that
   the momentum fraction carried by the spectator quark is:
   $\langle x\rangle \se 2 \left (1-M_Q^2/M_P^2\right )/3\,$
   ($M_P$ is the meson mass),\, and is $\se 15\%$ for the D meson and $\se
   6\%$ for the B one. So,\, the simplest estimate of $\langle {\ov
  \lambda}^2\rangle $ looks as: $\langle {\ov \lambda}^2\rangle/M_P^2\sim
   [1-2\langle x\rangle ]^2$ and is $\se 0.5$ for the D meson and $\se 0.8$
   for the B one.  More accurate calculations \cite{Cher1} give even stronger
   effect for the D meson:
   \bq \langle {\ov \lambda}^2\rangle_D\se 0.35\,M_D^2\,,\quad \langle{\ov
   \lambda}^2\rangle_B \se 0.80\,M_B^2\,.\eq
   As a result of all
   the above effects,\, the non-factorizable contributions become comparable
   with the factorizable one.

   \section{ $D^{\pm,o,s}$ decay widths.}

 Accounting for all the above described contributions,\, one obtains \cite
{Cher1}\,(in units $10^{-13}\,GeV$,\, the experimental values are given
in brackets,\, $\Gamma_{Born}=8.4$):
\bq \*\hspace{-0.3cm} \Gamma_{tot}(D^+)\se \Gamma_{Born}\left
[1.54-0.27-0.62-0.21+(0.26)_
{lept}\right ]= \tr = 0.70\, \Gamma_{Born}=5.9 \quad\{\,6.2\,\},\eq
\bq \Gamma_{tot}(D^o)\se \Gamma_{Born}\left [\,1.54-0.27+0.33+(0.26)_{lept}
\,\right ]=\tr = 1.86\, \Gamma_{Born}=15.6\quad   \{\,15.8\,\}\,,\eq
\bq \Gamma_{tot}(D^s)\se \Gamma_{Born}\left [\,1.54-0.27-0.07+(0.29)_{lept}
\,\right ]=\tr= 1.49\, \Gamma_{Born}=12.5 \quad  \{\,13.8\,\}\,.\eq
In Eqs.(31)-(33): the origin of "1.54" is explained in Eq.(21),\, and the
 term "-0.27" is the nonvalence contribution from figs.1a,1b diagrams
("-0.15" and "-0.12" respectively). The valence terms: "-0.62" in Eq.(31)
is the factorizable interference contribution (fig.2b without gluons);\,
"-0.21" in Eq.(31)
is the summary non-factorizable contribution of fig.2b and fig.3 diagrams;\,
"0.33" in Eq.(32) is the non-factorizable direct annihilation contribution,\,
fig.2a;\, the small term "-0.07" in Eq.(33) originates from the
Cabibbo-suppressed contributions and the non-factorizable annihilation
contributions.

The expected accuracy of $\Gamma_{tot}(D_i)$ calculations is not high: $\se
20\%$.
\footnote{
\,\,Besides,\, the SU(3) symmetry breaking corrections are
not accounted for in calculations with $D_s$.}
\,\, It is sufficient however to keep all the main contributions well under
control,\, and is much better in comparison with all previous estimates
which were qualitative rather than quantitative (see sect.7).

\section {$B^{\pm,o,s}$ lifetimes and $B-{\ov B}$ mixing.}

The ${\bar u}(d+s)-$ part of the $B_d$ width can be represented in the form:
\bq \Gamma_{(ud)}(B_d)\se \Gamma_{Born}\frac{(2C_+^2+C_-^2)}{3}\,
z_o^{ud}\,I_{rad}^{(ud)}\, \Omega_{power}\,,\eq
where: $(2C_+^2+C_-^2)/3\se 1.127$ is the renormalization factor due to the
evolution from $M_W$ to $M_b$\, \cite{Al},\, $\Omega_{power}\se 0.99$ is the
summary effect of all power corrections \cite{Cher1},\, $z_o^{ud}\se 0.46$ is
the phase space factor,\, and $I_{rad}^{(ud)}\se 1.03$ \cite{BBB} describes
all other $O(\alpha_s)$ corrections.  For the ${\bar c}(s+d)$ part:
$z_o^{cs}\se
 0.13,\, I_{rad}^{(cs)}\se 1.3$ \cite{Volosh},\,\cite{BBB}. On the whole (see
sect.11 for the semileptonic decays and $V_{cb}$,\,
$\Gamma_{Born}=G_F^2M_b^5|V_{cb}|^2/64\pi^3\se 3.93\cdot 10^{-13}\,GeV\,,
\,\, \tau_{Born}\se 1.67\,ps,$\, the data see in \cite{MD})\,:
\bq \Gamma_{tot}(B_d)\se \Gamma_{Born}\left
[\,0.53_{(ud)}+0.19_{(cs)}+ 2\cdot
0.114_{(e\nu+\mu\nu)}+0.028_{(\tau\nu)}+0.02_{(b\ra u)}\right ]\se \tr \se
 1.00\cdot \Gamma_{Born}\se 3.93\cdot 10^{-13}\,GeV\,,\quad
\Gamma_{tot}^{exper}(B_d)=(\,4.10\pm 0.18)\cdot 10^{-13}\,GeV\,,\eq
\bq Br_{(e\nu)}\se 11.4\%\,,\quad  Br_{(\tau\nu)}\se 2.8\%\,,
\quad Br_{(cs)}\se 19\%\,.\eq
All these results agree with data
\footnote{
\,\, One can expect also that higher loop corrections will increase
slightly the hadronic width.}
\,\,\,\, \cite{MD},\, although $Br_{(cs)}$ is slightly above the experimental
value: \,$Br^{(exper)}_{(cs)}=0.11\pm 0.06.$ \footnote{\,\, Let us
emphasize that $Br_{(cs)}$ will be essentially larger ($\geq 0.30$) for the
quark masses $M_c\se 1.4\,GeV,\, M_b\se 4.8\,GeV$,\, and this will constitute
a real difficulty.}

As for the lifetime differences,\, the qualitative picture remains the same as
for the D mesons but,\, of course,\, all the effects are much smaller.  Let us
denote by $\Gamma_o$ the common width of all $B^{\pm,o,s}$ mesons which they
have on neglect of the four-fermion operator contributions (and SU(3)
breaking). Then the valence contributions look as:
\bq \frac{\Delta \Gamma (B^+)}{\Gamma_o}\se (\,-2.1\%\,)+(\,-1.3\%\,)
\se -3.4\%\,,\eq
where $(-2.1\%)$ is the factorizable interference contribution (fig.2b
without gluons),\, and $(-1.3\%)$ is the summary non-factorizable
contribution of fig.2b and fig.3 diagrams.

The $B_d$ and $B_s$ mesons (neglecting SU(3) breaking) receive only
the non-factorizable contributions from fig.2a and fig.3 diagrams:
\bq \frac{\Delta {\ov \Gamma}(B_d)}{\Gamma_o}\se 0.6\%\,,\quad \frac{\Delta
{\ov \Gamma} (B_s)}{\Gamma_o}\se 0.5\%\,.\eq
Finally,\, the four-fermion operators give also the non-valence
contributions,\,
both factorizable: $\se (-0.6\%)$ from fig.1a diagram,\, and non-
factorizable: $\se (-0.1\%)$ from fig.1b one.

On the whole,\, the lifetime difference of the $B_d$ and $B^{\pm}$
mesons is \cite{Cher1}\,:
\bq \frac{{\ov \Gamma}(B_d)-\Gamma (B^{\pm})}{{\bar \Gamma}(B)}\se 4\%,\,\eq
while $\Delta {\ov \Gamma}$ of $B_d$ and $B_s$ is zero within the available
accuracy.

 The above described non-factorizable contributions,\, figs.2,\,3,\, determine
also the corrections to the factorization approximation for the ${\ov B}-
B$-mixing,\, and appear to be surprisingly large here \cite{Cher1}:
$\se -18\%\,.$ As a result,\, one obtains for the $B_s$-mesons (neglecting
SU(3)-breaking):
\bq \frac{\Gamma(B_s^{short})-\Gamma(B_s^{long})}{{\ov \Gamma}(B_s)}\se
6\%\,,\eq
and for the "bag factor" of the ${\ov B}_d-B_d$ mass mixing:
\bq <{\ov B}_d|({\ov b}\,\Gamma_\nu\,d)({\ov b}\,
\Gamma_\nu\,d)|B_d>_{\mu=M_b}\equiv
\frac{8}{3}f^2_B(M_b) M_B^2 B_B(M_b)\,, \nonumber \eq
\bq B_B(M_b)\se (1-18\%)= 0.82\,.\eq

\section {$B\ra e\,\nu+X_c$:\, determination of $|V_{cb}|$.}

The calculation proceeds in analogy with those in sect.4.\, The total effect
of $O( \Lambda_{QCD}^2/M_b^2)$ power corrections is much smaller now ($\se
-4\%$),\, and $\delta^{(b)}_{lept}$ is negligibly small ($\se -2\cdot
10^{-3}$)\, \cite{Cher1}.
Most important are radiative corrections which look as \cite
{LSW} ($f_o\se 2.46$ here):
\bq I_{rad}\se\left [1-\frac{2}{3}\frac{\alpha_s(M_b)}{\pi}f_o-1.68\,b_o\left
( \frac{\alpha_s(M_b)}{\pi}\right )^2-\cdots \right ]\se \tr \se
\left [1-\frac{2}{3}
\frac{\alpha_s(M_b)}{\pi}f_o\left (1+0.6+\cdots\right )\right ]=\left [
1-\frac{2}{3}\frac{\alpha_s(M_b)}{\pi}f_o\,\kappa_b^{(w)}\right ]\,.\eq
Comparing with Eq.(3) it is seen that (analogously to the charm case)
two series follow the same pattern. So,\, we estimate:
\,$\kappa_b^{(w)}\se 2.1,\, I_{rad}\se 0.776\,.$ Therefore ($z_o^{
e\nu}\se 0.46$):
\bq \Gamma (B\ra e\,\nu+X_c)\se 0.44\,\Gamma_{Born}^{sl}\,I_{rad}\se 4.5
\cdot 10^{-14}\,GeV \left |\frac{V_{cb}}{0.042}\right |^2\,, \eq
\bq \Gamma_{Born}^{sl}=\frac{G_F^2\,M_b^5|V_{cb}|^2}{192\,\pi^3}\se 13.1\cdot
10^{-14}\,GeV\left |\frac{V_{cb}}{0.042}\right |^2\,.\eq
As for the data,\, we take \cite{MD}:
\bq \tau_B=(\,1.60\pm 0.07\,)\,ps\,,\quad Br(B\ra e\nu+X)=(11.0\pm
0.5)\%\,, \eq
and obtain:
\bq |V_{cb}|=(\,42\pm 1\,)\cdot 10^{-3}\left [\frac{Br(B\ra e\nu+X)}
{11.0\%} \right ]^{1/2}\left [\frac{1.6\,ps}{\tau_B}\right ]^{1/2}\,.\eq
The error bars in Eq.(46) were estimated by varying: $1.60\leq M_c \leq
1.70\,GeV,\\ 3.37\leq (M_b-M_c)\leq 3.41\,GeV\,.$
\footnote{
\,\,For comparison,\, using $M_c=1.4\,GeV,\,M_b=4.8\,GeV\,$ and
proceeding in the same way one obtains 44.5 instead of 42.0 in Eq.(46),\, and
it seems it will be difficult to reconcile this value with the data on
$\Gamma(B\ra D^*e\nu)$.}

For the $B\ra \tau\nu+X$ decays: $f_o\se 2.0,\, z_o^{\tau\nu}\se 0.105,\,
I_{rad}\se 0.83,\,$ and so:
\bq Br\left (\frac{\tau}{e}\right )\se 0.25\,.\eq

\section {The unitarity triangle.}

{\bf 1)\,.}  Using: $V_{cb}=A\,\lambda^2\,,\quad \lambda\se 0.221$\,\,
and (see Eq.(46))\, $V_{cb}\se 4.2\cdot 10^{-2}$\,, one has:
\bq A \se 0.86\,. \eq

{\bf 2)\,.} The $B^o-{\bar B}^o$ mass difference is given by the well known
formula (see i.e. \cite{BH}),\, and using ( see Eqs.(20),\,(41)\, ):
$f_B(M_b)=120\,MeV,\, B_B(M_b)=0.82,\,$ and \cite{MD}:
\bq  M_t^{pole}=175\,GeV, \quad \tau(B_d)=1.6\,ps\,,\eq
one obtains :
\bq x_d=\frac{\Delta M_d}{\Gamma(B_d)}=(0.78\pm 0.06)
\se 4.5\cdot 10^3\left |V_{td}\right |^2\,,\eq
\bq \left |V_{td}\right |=\left |V_{cb}\right |\lambda \left [(1-\rho)^2+
\eta^2\right ]^{1/2}\se 1.3\cdot 10^{-2}\,,\eq
\bq \left [\,(1-\rho)^2+\eta^2\right]\se 2.0\,.\eq

{\bf 3)\,.} Using the above given parameters and ${\hat B}_K\se 0.82$ from
the lattice calculations \cite{Gupta},\,
the CP-violating part of the $K^{o}-{\bar K}^
{o}$ mixing can be written in the form (see i.e. \cite{BH}):
\bq e^{-i\pi/4}\epsilon_K\cdot 10^3\se 8.0\,{\hat B}_K\,\eta\,
(\,1.36-\rho\,)=2.26\,, \nonumber \eq
\bq \eta\,(\,1.36-\rho\,)\se 0.35\,.\eq
Therefore,\, we obtain from Eqs.(52),\,(53):
\bq \rho\se -0.40\,,\quad \eta\se 0.20\,,\quad \delta=arctg\,(\frac{\eta}
{\rho})\se 0.85\,\pi\,,\quad \left |\frac{V_{ub}}
{V_{cb}}\right |\se 0.10\,, \eq
\bq \sin\,2\alpha\se 0.60,\,\quad \sin\,2\beta\se 0.28,\,\quad
\sin\,2\gamma\se -0.80\,.\eq
The unitarity triangle is shown in fig.4\,.
With the above parameters the
CP-violating asymmetry in the $B^o_d\ra\Psi\,K_S$ decay is:
\bq |A(B^o_d\ra \Psi\,K_S)|\se \frac{x_d}{1+x_d^2}
\sin\, 2\beta\se 0.14\,.\eq


\section {\bf Summary}

It is seen from all the above presented calculations that a
self-consistent picture emerges which agrees with a large number of
various experimental data,\, and allows to obtain a number of important
predictions which can be checked in future experiments.

One of our main concerns was to calculate reliably the four-fermion operator
contributions. These are of crucial importance for explaining the pattern of
the $D^{\pm,o,s}$ lifetimes,\, the $B^{\pm,o,s}$ lifetime differences and the
${\ov B}-B$ mixing. There is a clear reason explaining the importance of these
four-fermion operator contributions,\, although they are formally only
$O(\Lambda^3_{QCD}/M_Q^3)$ corrections: they are the first who gain the large
numerical factor $\se 16\pi^2$  (see Eq.(25)) due
to the two-particle phase space,\, in
comparison with the three-particle one for the Born term and (non-valence)
$O(\Lambda^2_{QCD}/M_Q^2)$ corrections. It is clear that this enhancement
factor operates one time only,\, so that all other $O(\Lambda^3_{QCD}/M_Q^3)$
and higher order corrections are naturally small (see Eq.(14)) and have no
much chances to be of real importance. \footnote{\, The applicability of the
standard operator expansions to the calculation of $\Gamma(D\ra e\nu+X)$ has
been questioned in \cite {BDS} on the only ground that the authors don't
believe the c-quark pole mass may be as large as $M_c\se 1.65\,GeV$. They
insist it can not exceed $\se 1.4\,GeV$. Let us emphasize that the calculated
value of $\Gamma(D\ra e\nu+X)$ (see Eq.(15)) will decrease $\se 3$ times for
$M_c=1.4\,GeV$.  So,\, there should be huge additional contributions which
dominate the semileptonic width and remain invisible within the standard
operator expansion. No one reliable argument is presented however in \cite
{BDS} to justify $M_c\se 1.4\,GeV$,\, and no one missed contribution is shown
which is of great importance.}

The calculated values of the ${\bf c}$ and ${\bf b}$ quark pole masses,\,
$M_c\se 1.65\,GeV,\, M_b\se 5.04\,GeV,\,$
appeared to be significantly larger the widely accepted at present
values: $M_c\se 1.35-1.45\,GeV,\, M_b\se 4.8\,GeV$. This difference is of
great importance,\, as most of calculations with heavy quarks are highly
sensitive to precise values of their masses. In particular,\, the calculated
decay constant $f_B\se 120\,MeV$ appeared to be much smaller the widely
accepted at present value $f_B\se 180-200\,MeV$. This difference leads,\, in
its turn,\, to essentially different predictions for the $B^o-B^{\pm}$
lifetime difference,\, ${\ov B}^o-B^o$ mixing and the unitarity triangle
parameters.  Just because the above obtained values of $M_c,\, M_b,\, {\ov
\Lambda},\, f_B,\,$
etc,\, look highly non-standard at present,\, we described above the
calculations of many experimentally measured quantities to show there is no
disagreement with data. Besides,\, a number of concrete predictions is
described which can be checked in future experiments (see Tables).\\

\begin{center} \hspace{-2cm}{\bf ACKNOWLEDGEMENT}\end{center}

I am deeply indebted to organizers of this useful and interesting seminar,\,\\
Prof. J.K{\"o}rner and Prof. P.Kroll,\, for a kind invitation and support.\\

\newpage
\begin{center}\hspace {-3cm} {\bf REFERENCES} \end{center}


\newpage
\vspace{-1 cm}
\begin{center} Table 1: {\bf INPUT}  \end{center}
\vspace {0.5cm}
\begin{center}
\begin{tabular}{|c|c|}
\hline
$\Gamma(D\ra e\nu+X)$ & $(1.08\pm 0.06)\cdot 10^{-13}\,GeV$ \\ \hline
$\tau(B_d)$ & $(1.60\pm 0.07)\, ps $ \\ \hline
$Br(B\ra e\nu+X)$ & $(11.0\pm 0.5)\%  $ \\ \hline
$x_d=\Delta m_d/\Gamma_{B_d}$ & $0.78\pm 0.06 $ \\ \hline
$\epsilon_K$ & $2.26\cdot 10^{-3}$ \\ \hline
${\hat B}_K$ & $0.82\pm 0.05 $  \\ \hline
$M_t^{pole}$ & $175\, GeV$ \\ \hline
$\lambda=|V_{us}|$ & $0.221$  \\ \hline
$\alpha_{\overline {MS}}(M_W)$ & 0.118  \\ \hline
$[f_B^2 B_B]_{B_s}/[f_B^2 B_B]_{B_d}$ & 1.3 \\ \hline
\end{tabular}
\end{center}
\vspace{1 cm}

\begin{center} Table 2: {\bf OUTPUT} \end{center}
\vspace{0.5cm}
\begin{center}
\begin{tabular}{|c|c|}
\hline
$M_c^{pole}$ & $1.65\,$ GeV \\ \hline
$M_b^{pole}$ & $5.04\,$ GeV \\ \hline
$f_D(M_c)$ & $165\,$ MeV \\ \hline
$f_B(M_b)$ & $120\,$ MeV \\ \hline
$B_B(M_b)$ & 0.82 \\ \hline
$\left [\,{\ov \Gamma}(B_d)-\Gamma(B^{\pm})\,\right ]/
\Gamma_B$ & $4\%$ \\ \hline
$\left [\,\Gamma(B_s^{short})-\Gamma(B_s^{long})\,\right ]/ {\ov
\Gamma} (B_s)$ & $8\%$ \\ \hline
$\left [\,\Gamma(B_d^{short})-\Gamma(B_d^{long})\,\right ]/
{\ov \Gamma} (B_d)$ & $0.6\%$ \\ \hline
$\Delta m_s/\Delta m_d$ & 14 \\ \hline
$|V_{cb}|$ & $4.2\cdot 10^{-2}$ \\ \hline
$|V_{td}|$ & $1.3\cdot 10^{-2}$ \\ \hline
$|V_{ub}/V_{cb}|$ & 0.10 \\ \hline
$|A(B\ra \Psi\,K_s)|$ & 0.14 \\ \hline
$|\epsilon^{\prime}/\epsilon|\cdot 10^4$ & a few units \\ \hline
\end{tabular}
\end{center}

\vspace{1cm}

\hspace{2 cm} {\bf Figure captions}\\

Fig.1 $\quad$ Non-valence factorizable (a,c) and non-factorizable
(b) contributions to matrix elements.

Fig.2 $\quad$ Valence direct (a) and cross
(b) annihilation non-factorizable contributions to matrix
elements of coloured operators; (c) the same for colourless operators.

Fig.3 $\quad$ Valence non-factorizable contribution (plus the
mirror diagram).

Fig.4 $\quad$ The unitarity triangle: $\rho\se -0.40\,,\,\,\,
\eta\se 0.20\,.$

\end{document}